\documentclass[a4]{article}

\usepackage{amsmath}
\usepackage{amsfonts}

\newcommand{\ol}{\overline}
\newcommand{\wt}{\widetilde}

\def\tr{\mathop{\rm tr}\nolimits}
\def\pexp{\mathop{\rm Pexp}\nolimits}
\def\rank{\mathop{\rm rank}\nolimits}

\newcommand{\RR}{\mathbb{R}}
\newcommand{\TT}{\mathbb{T}}

\begin{document}
\begin{titlepage}
\title{
\begin{flushright}
\normalsize{TIT/HEP-612\\
Apr 2011}
\end{flushright}
       \vspace{2cm}
Relation between the 4d superconformal index and the $S^3$ partition function
       \vspace{2cm}}
\author{
Yosuke Imamura\thanks{E-mail: \tt imamura@phys.titech.ac.jp}$^{~1}$
\\[30pt]
{\it $^1$ Department of Physics, Tokyo Institute of Technology,}\\
{\it Tokyo 152-8551, Japan}
}
\date{}

\maketitle
\thispagestyle{empty}

\vspace{0cm}

\begin{abstract}
\normalsize
A relation between the 4d superconformal index and the ${\bf S}^3$ partition function
is studied with focus on the 4d and 3d actions used in localization.
In the case of vanishing Chern-Simons levels and round
${\bf S}^3$
we explicitly show that the 3d action
is obtained
from the 4d action
by dimensional reduction
up to terms which do not affect the exact results.
By combining this fact and a recent proposal concerning
a squashing of ${\bf S}^3$
and $SU(2)$ Wilson line, we obtain a formula
which gives the partition function
depending on the Weyl weight of chiral multiplets,
real mass parameters, FI parameters, and a squashing parameter
as a limit of the index of a parent 4d theory.
\end{abstract}

\end{titlepage}


\section{Introduction}\label{intro.sec}
Recent years, exactly calculable quantities
in gauge theories play important roles
in study of gauge theories themselves and their relation to string/M theory.
In this paper we discuss a relation between two of such quantities.

One is
the ${\bf S}^3$ partition function\cite{Kapustin:2009kz,Jafferis:2010un,Hama:2010av}.
It is used to confirm dualities among 3d theories
\cite{Kapustin:2010xq,Jafferis:2011ns,Kapustin:2011gh,Willett:2011gp}
and predictions of AdS$_4$/CFT$_3$
\cite{Drukker:2010nc,Herzog:2010hf,Martelli:2011qj,Cheon:2011vi,Jafferis:2011zi}.
Furthermore, this function provides a simple way to determine
the R-charge at IR fixed points\cite{Jafferis:2010un}.
The partition function is evaluated exactly
by localization.
We choose a nilpotent supercharge ${\cal Q}$ and
deform the action by ${\cal Q}$-exact terms.
The partition function is given by
\begin{equation}
Z=\int {\cal D}\Phi \exp\left(
-S_0^{(3d)}-u\int_{{\bf S}^3}\sqrt{g}{\cal L}^{(3d)} d^3x
\right).
\label{zs3}
\end{equation}
$S^{(3d)}_0$ is the original action of the 3d theory and
the second term in the exponent is the ${\cal Q}$ exact action.
This path integral does not depend on $u$,
and is evaluated exactly in the weak coupling limit $u\rightarrow\infty$.

The other exactly calculable quantity we consider is
the ${\cal N}=1$ superconformal index
for 4d theories\cite{Kinney:2005ej,Romelsberger:2005eg}.
The index is defined by
\begin{equation}
I(t,x,h_i)=\tr\left[(-1)^F
q^{D-\frac{3}{2}R-2J_L}
t^{R+2J_L}
x^{2J_R}
\prod_ih_i^{{\cal F}_i}\right],
\label{indexdef}
\end{equation}
where $D$ (the dilatation), $R$ (the R-charge),
$J_L$ and $J_R$ ($SU(2)_L\times SU(2)_R$ spins),
are Cartan generators of the ${\cal N}=1$ superconformal algebra
$PSU(2,2|1)$,
and ${\cal F}_i$ are Cartan generators of the flavor symmetry.
Only operators saturating the BPS bound
\begin{equation}
D-\frac{3}{2}R-2J_L\geq 0
\end{equation}
contribute to the index, and
(\ref{indexdef}) is independent of the
variable $q$.
This quantity is exactly calculable, and is conveniently used as a tool to check
Seiberg-duality\cite{Romelsberger:2007ec,Dolan:2008qi,Spiridonov:2008zr,Spiridonov:2009za}
and AdS$_5$/CFT$_4$\cite{Kinney:2005ej,Nakayama:2005mf,Nakayama:2006ur,Benvenuti:2006qr,Gadde:2010en}.

One way to compute the index is to use localization.
We choose a nilpotent supercharge ${\cal Q}$ and
deform the action by $\cal Q$-exact terms.
The index can be expressed in the path integral form
\begin{equation}
I(t,x,h_i)=\int {\cal D}\Phi \exp\left(
-S_0^{(4d)}-u\int_{{\bf S}^3\times{\bf S}^1}\sqrt{g}{\cal L}^{(4d)} d^4x
\right),
\label{ideform}
\end{equation}
where $S_0^{(4d)}$ is the original action of the theory
defined in ${\bf S}^3\times{\bf S}^1$,
and the second term in the exponent is the ${\cal Q}$-exact deformation action.
The chemical potentials are introduced as non-trivial Wilson lines around
${\bf S}^1$.
Let $r$ and $\beta r$ be the ${\bf S}^3$ radius and the ${\bf S}^1$ period,
respectively.
The ratio $\beta$ is related to the parameter $q$ by
$q=e^{-\beta}$.
In the case of the index, the deformation term does not necessarily
have to be ${\cal Q}$-exact
because the index does not depend on continuous coupling constants;
even so, we adopt a ${\cal Q}$-exact deformation action in this paper
for the reason which will become clear shortly.

The similarity between (\ref{zs3}) and (\ref{ideform}) 
strongly suggests that there exists some relation between
the index and the partition function.
If we consider 4d and 3d theories with the same gauge group $G$
and the same matter contents,
we naturally expect
that the partition function is obtained by taking a small ${\bf S}^1$ limit
of the index.
Such a relation was recently studied in \cite{Dolan:2011rp,Gadde:2011ia}.

In \cite{Dolan:2011rp} it is shown for particular examples of gauge theories
that a relation between the 3d partition function and the 4d index follows
from certain mathematical properties of special functions appearing in the
index and partition function.
A similar relation is also studied in
\cite{Gadde:2011ia}, and a limiting procedure which reduces
the superconformal index of 4d ${\cal N}=2$ theories
to the ${\bf S}^3$ partition function of corresponding
3d ${\cal N}=4$ theories is proposed.
In these works, only the final expressions for
the partition function and the index are studied,
and physical origin of the relation is not so obvious.
The purpose of this paper is to extend the relation obtained
in \cite{Dolan:2011rp,Gadde:2011ia} to
general 3d ${\cal N}=2$ and 4d ${\cal N}=1$ theories,
and to establish the relation
at more fundamental level
by comparing 3d and 4d actions.
For this purpose, it is convenient to use as similar deformation terms
as possible in two computations.
We use ${\cal Q}$-exact deformation terms in both cases with
closely related supercharges ${\cal Q}$ in 3d and 4d theories.

In both (\ref{zs3}) and (\ref{ideform}),
the deformation terms dominate the actions
in the weak coupling limit $u\rightarrow\infty$,
and only few terms in the original actions are relevant to the partition function and the index.
Let $S^{(3d)}_{\rm rel}$ and $S^{(3d)}_{\rm rel}$
be the relevant terms including the deformation terms.
$S^{(3d)}_{\rm rel}$ consists of
(supersymmetric completion of) Chern-Simons and FI terms
in the original action $S^{(3d)}_0$ and
the ${\cal Q}$-exact terms
\begin{equation}
S^{(3d)}_{\rm rel}=S^{(3d)}_{\rm CS}+S_{\rm FI}^{(3d)}+
u\int_{{\bf S}^3}\sqrt{g}{\cal L}^{(3d)}d^3x,
\label{relevant}
\end{equation}
while $S_{\rm rel}^{(4d)}$ consists of
the (supersymmetric completion of) FI terms
and the deformation terms
\begin{equation}
S^{(4d)}_{\rm rel}=S_{\rm FI}^{(4d)}+
u\int_{{\bf S}^3\times{\bf S}^1}\sqrt{g}{\cal L}^{(4d)}d^4x.
\label{relevant4d}
\end{equation}
We consider 3d and 4d theories with the same gauge group $G$ and
chiral multiplets $\Phi_I$
belonging to the same $G$-representations $R_I$.
We assume that the Weyl weight $\Delta_I$\footnote{%
The deformation terms are not invariant under the
dilatation, and the dilatation is broken in the deformed theories.
For this reason, the parameters $\Delta_I$ in the deformed theories
should be regarded
not as the weyl weights
but as parameters appearing in the ${\cal Q}$ transformation
laws for chiral multiplets.
The absence of the dilatation symmetry in the deformed theories
does not cause any problem because
we need only the fermionic symmetry ${\cal Q}$
for the computation of the exact results.
} of each chiral multiplet
is the same in 3d and 4d.
We explicitly show
for a 3d theory without Chern-Simons terms on round ${\bf S}^3$
that $S_{\rm rel}^{(3d)}$ is obtained
by dimensional reduction
of $S_{\rm rel}^{(4d)}$
provided that an appropriate Wilson line is turned on.

The symmetry associated with the Wilson line may not be a
symmetry of the original action $S_0^{(4d)}$,
but is a symmetry of $S^{(4d)}_{\rm rel}$.
(The symmetry may be anomalous.
We discuss the treatment of anomalous symmetries
at the end of \S\ref{comparison.sec}.)
$S^{(4d)}_{\rm rel}$ has the symmetry rotating each chiral multiplet
independently.
For each chiral multiplet $\Phi_I$
we define the charge ${\cal F}_I$ rotating only $\Phi_I$
and the corresponding chemical potential $h_I$.
By comparing $S_{\rm rel}^{(3d)}$
and $S_{\rm rel}^{(4d)}$,
we obtain a formula which
gives the partition function $Z$
as a small radius limit of the index
$I(t,x,h_I)$.
We further generalize the relation
by using the recently proposed\cite{Gadde:2011ia} connection
between squashing parameter $s$ of ${\bf S}^3$\cite{Hama:2011ea}
and $SU(2)_R$ Wilson line.
The most general formula we propose in this paper is
\begin{equation}
Z=\lim_{q\rightarrow 1}I(t=q,x=q^s,h_I=q^{-ir\mu_I+\frac{1}{3}\Delta_I})
|_{\zeta_A^{(4d)}=\frac{1}{\beta r}\zeta_A^{(3d)}},
\label{mainresult}
\end{equation}
where $\mu_I$ are real mass parameters, and
$\zeta_A^{(4d)}$ and $\zeta_A^{(3d)}$ are 4d and 3d FI parameters, respectively.
Unfortunately, when Chern-Simons levels $k_a$ of 3d theory
are non-vanishing,
we could not reproduce the ${\bf S}^3$
partition function from the index due to the difficulty in obtaining
Chern-Simons terms
by dimensional reduction.

The paper is organized as follows.
After explaining our notation for spinors in the next section,
we summarize the superconformal algebra and the supersymmetry
transformation laws in \S\ref{algebra.sec} and \S\ref{susy.sec}.
Exact computations of the ${\bf S}^3$ partition function and
the 4d superconformal index are briefly reviewed in
\S\ref{partition.sec} and \S\ref{index.sec},
respectively.
In \S\ref{comparison.sec} we compare the 3d and 4d actions,
and find the relation
between the partition function and the index in the case
of $\mu_I=k_a=\zeta^{(3d)}_A=s=0$.
Generalization to non-vanishing parameters is discussed in \S\ref{generalization.sec}.
Conclusions are presented in \S\ref{conc.sec}.

\section{Notation for spinors}\label{notation.sec}
Because we consider both 3d and 4d theories,
we use notation for spinors such that the expression of 3d
and 4d theories look as similar as possible.

For 3d spacetime,
we use coordinates $x^m$ ($m=1,2,3$).
Although we can define Majorana spinors in
3d Minkowski spacetime, all spinors we use
are complex spinors.
For a complex spinor $\psi$, we denote its Majorana conjugate by $\ol\psi$.
In Euclidean spacetime $\psi$ and $\ol\psi$ should be treated as
independent spinors.

For 4d spacetime, we use coordinates
$x^\mu$ ($\mu=1,2,3,4$).
When we consider ${\bf S}^3\times{\bf S}^1$ background,
we use $x^m$ for ${\bf S}^3$ and $x^4$ for ${\bf S}^1$.
The 4d Dirac's matrices are expressed
in terms of the 3d Dirac's matrices by
\begin{equation}
\gamma^m=\left(\begin{array}{cc} 0 & \gamma^m \\
\gamma^m & 0 \end{array}\right)\quad m=1,2,3,\quad
\gamma^4=\left(\begin{array}{cc}
0 & -i \\
i & 0
\end{array}\right).
\end{equation}
We use the same symbol $\gamma^m$ for 3d and 4d Dirac's matrices.
The charge conjugation and the chirality in 4d are
\begin{equation}
C_{ab}=\left(\begin{array}{cc} \epsilon_{ab} & 0 \\
0 & \epsilon_{ab} \end{array}\right),\quad
\gamma^5=\left(\begin{array}{cc} {\bf 1}_2 & 0 \\
0 & -{\bf 1}_2 \end{array}\right).
\end{equation}
We call the upper (lower) half of a four-component spinor
left-handed (right-handed).
Namely, a left-handed (right-handed) spinor has positive (negative) chirality.
3d and 4d completely anti-symmetric tensors $\epsilon_{mnp}$ and $\epsilon_{\mu\nu\rho\sigma}$ are defined by
\begin{equation}
\gamma_{mnp}=i\epsilon_{mnp}{\bf1}_2,\quad
\gamma^5\gamma_{\mu\nu\rho\sigma}=-\epsilon_{\mu\nu\rho\sigma}{\bf1}_4.
\end{equation}
We raise and lower spinor indices by
the relation $\psi_a=\psi^b\epsilon_{ba}$.
Spinor indices are contracted by the
NW-SE rule.
For example, for spinors $\psi$ and $\chi$,
$\psi\chi\equiv\psi^a\chi_a=\psi^a\chi^b\epsilon_{ba}$.

In 4d we use two-component representation.
We use a symbol without and with bar for a left-handed and right-handed
spinor, respectively.
For example, when we use symbol $\psi$ and $\ol\psi$ as 4d
two-component spinors,
their four-component representations are
\begin{equation}
\left(\begin{array}{c}\psi \\ 0\end{array}\right),\quad
\left(\begin{array}{c}0 \\ \ol\psi\end{array}\right).
\end{equation}
Note that $\ol\psi$ is not the Dirac's conjugate of $\psi$.
We will never use Dirac's conjugate in this paper.

We use indices $\mu,\nu,\ldots$ not only in 4d but also in 3d.
In that case
we assume that all fields do not depend on $x^4$,
and the $4$-th component of a gauge field $A_\mu$ is regarded as a
Hermitian scalar field $\sigma$.
For example, if the gauge covariant derivative is given by
$D_\mu=\partial_\mu-iA_\mu$,
the fermion kinetic term $-(\ol\psi\gamma^\mu D_\mu\psi)$
represents in 3d the sum of two terms
$-(\ol\psi\gamma^m D_m\psi)$ and 
$-(\ol\psi\sigma\psi)$.

\section{Superconformal algebra}\label{algebra.sec}
Before considering actions and transformation laws,
let us compare the 4d ${\cal N}=1$ superconformal algebra
and 3d ${\cal N}=2$ superconformal algebra.

The 4d algebra contains the generators
\begin{equation}
M_{\mu\nu},\quad
P_\mu,\quad
K_\mu,\quad
D,\quad
R,\quad
Q,\quad
\ol Q,\quad
S,\quad
\ol S,
\end{equation}
while the 3d algebra contains
the same generators with vector indices $\mu$ and $\nu$ running over
$1,2,3$ only.
For later use we define Cartan generators of the rotation groups,
\begin{equation}
M_{12}=iJ_3\quad(3d),\quad
M_{12}=i(J_L+J_R),\quad
M_{34}=i(J_L-J_R)\quad(4d).
\end{equation}

Almost all (anti-)commutation relations are the same in 3d and 4d.
\begin{align}
&[M_{\mu\nu},M_{\rho\sigma}]=
\eta_{\mu\rho}M_{\nu\sigma}
-\eta_{\mu\sigma}M_{\nu\rho}
-\eta_{\nu\rho}M_{\mu\sigma}
+\eta_{\nu\sigma}M_{\mu\rho},\nonumber\\
&[M_{\mu\nu},P_\rho]=\eta_{\mu\rho}P_\nu-\eta_{\nu\rho}P_\mu,\quad
[M_{\mu\nu},K_\rho]=\eta_{\mu\rho}K_\nu-\eta_{\nu\rho}P_\mu,\nonumber\\
&[D,P_\mu]=P_\mu,\quad
[D,K_\mu]=-K_\mu,\quad
[P_\mu,K_\nu]=-2M_{\mu\nu}+2\eta_{\mu\nu}D,
\nonumber\\
&[M_{\mu\nu},{\cal S}]=\frac{1}{2}\gamma_{\mu\nu}{\cal S},\quad
({\cal S}=Q,\ol Q,S,\ol S),\nonumber\\
&[R,Q]=-Q,\quad
[R,\ol Q]=\ol Q, \quad
[R,S]=S,\quad
[R,\ol S]=-\ol S,
\nonumber\\&
[D,Q]=\frac{1}{2}Q,\quad
[D,\ol Q]=\frac{1}{2}\ol Q, \quad
[D,S]=-\frac{1}{2}S, \quad
[D,\ol S]=-\frac{1}{2}\ol S,
\nonumber\\&
[S,P_\mu]=\gamma_\mu\ol Q, \quad
[\ol S,P_\mu]=\gamma_\mu Q, \quad
[Q,K_\mu]=\gamma_\mu\ol S, \quad
[\ol Q,K_\mu]=\gamma_\mu S,
\nonumber\\&
\{Q_a,\ol Q_b\}=2(\gamma^\mu)_{ab}P_\mu, \quad
\{S_a,\ol S_b\}=2(\gamma^\mu)_{ab}K_\mu.
\label{common}
\end{align}
Differences between 3d and 4d arise only in
$\{S,Q\}$ and $\{\ol S,\ol Q\}$.
In the 3d algebra, they are
\begin{align}
\{S_a,Q_b\}=&(\gamma^{mn})_{ab}M_{mn}+2\epsilon_{ab}D+2\epsilon_{ab}R,
\nonumber\\
\{\ol S_a,\ol Q_b\}=&(\gamma^{mn})_{ab}M_{mn}+2\epsilon_{ab}D-2\epsilon_{ab}R,
\label{qs3d}
\end{align}
while in the 4d algebra,
the coefficients of the $R$-charge terms are different.
\begin{align}
\{S_a,Q_b\}=&(\gamma^{\mu\nu})_{ab}M_{\mu\nu}+2\epsilon_{ab}D+3\epsilon_{ab}R,
\nonumber\\
\{\ol S_a,\ol Q_b\}=&(\gamma^{\mu\nu})_{ab}M_{\mu\nu}+2\epsilon_{ab}D-3\epsilon_{ab}R.
\label{qs4d}
\end{align}
In radial quantization, the dilatation $D$ is regarded as Hamiltonian,
and $\ol Q^a$ and $\ol S_a$ are treated to be Hermitian conjugate to each other.
From (\ref{qs3d}) and (\ref{qs4d}) we can derive BPS bounds.
In particular, the bound obtained from $\{\ol S_1,\ol Q^1\}$
is important in the following computations.
In 3d, it is
\begin{equation}
\{\ol S_1,\ol Q^1\}=2D-2R-2J_3\geq 0.
\end{equation}
In 4d, we obtain the bound with different coefficients
\begin{equation}
\{\ol S_1,\ol Q^1\}=2D-3R-4J_L\geq 0.
\end{equation}

\section{Supersymmetry transformations}\label{susy.sec}
Because the Poincare subalgebra
in (\ref{common})
generated by
$M_{\mu\nu}$, $P_\mu$, $Q$ and $\ol Q$ is the same in 3d and 4d,
(up to the absence of $P_4$ and $M_{m4}$ in 3d,)
$Q$ and $\ol Q$-transformation laws in the flat background
take the same form in 3d and 4d.
For a vector multiplet
$(A_\mu,\lambda,\ol\lambda,D)$,
the $Q$ and $\ol Q$-transformations are
\begin{align}
\delta^0 A_\mu=&i(\epsilon\gamma_\mu\ol\lambda)-i(\ol\epsilon\gamma_\mu\lambda),\nonumber\\
\delta^0\lambda=&
\frac{i}{2}\gamma^{\mu\nu}\epsilon F_{\mu\nu}
+D\epsilon,\nonumber\\
\delta^0\ol\lambda=&
-\frac{i}{2}\gamma^{\mu\nu}\ol\epsilon F_{\mu\nu}
+D\ol\epsilon,\nonumber\\
\delta^0D=&
-(\epsilon\gamma^\mu D_\mu\ol\lambda)-(\ol\epsilon\gamma^\mu D_\mu\lambda).
\label{d0vector}
\end{align}
We use the symbol $\delta^0$ rather than $\delta$ to
emphasize that these are rules for the flat background.
When we regard these as rules for 3d theory,
all fields are assumed to be independent of $x^4$, and
$A_4$ should be regarded as a Hermitian scalar field $\sigma$.
For a chiral multiplet $(\phi,\psi,F)$,
the transformation laws are
\begin{align}
\delta^0 \phi=&\sqrt2(\epsilon\psi),\nonumber\\
\delta^0 \phi^\dagger=&\sqrt2(\ol\epsilon\ol\psi),\nonumber\\
\delta^0\psi=&
-\sqrt{2}\gamma^\mu\ol\epsilon D_\mu\phi
+\sqrt{2}\epsilon F,\nonumber\\
\delta^0\ol\psi=&
-\sqrt{2}\gamma^\mu\epsilon D_\mu\phi^\dagger
+\sqrt{2}\ol\epsilon F^\dagger,\nonumber\\
\delta^0 F=&
-\sqrt{2}(\ol\epsilon\gamma^\mu D_\mu\psi)
-2(\ol\epsilon\ol\lambda)\phi,\nonumber\\
\delta^0 F^\dagger=&
-\sqrt{2}(\epsilon\gamma^\mu D_\mu\ol\psi)
-2\phi^\dagger(\epsilon\lambda).
\label{d0chiral}
\end{align}

We can construct supersymmetry transformation laws for
an arbitrary conformally flat background
from (\ref{d0vector}) and (\ref{d0chiral})
by Weyl-covariantization.
By a Weyl transformation
\begin{equation}
e_\mu^a=e^{-\alpha}e_\mu'^a,
\label{weyle}
\end{equation}
a field $\varphi$ with Weyl weight $\Delta_\varphi$ is transformed by
\begin{equation}
\varphi=e^{\Delta_\varphi\alpha}\varphi'.
\label{weylphi}
\end{equation}
Even if a field $\varphi$ has definite Weyl weight,
its derivative is not transformed covariantly as (\ref{weylphi})
and terms containing $\partial_\mu\alpha$ arise.
There are such non-covariant terms in the transformation laws
(\ref{d0vector}) and (\ref{d0chiral}).
To extend them to a general conformally flat background,
we should covariantize them with respect to Weyl transformation
by adding terms containing
derivatives of parameters $\epsilon$ and $\ol\epsilon$.
$\delta^0\lambda$ in 3d and $\delta^0\psi$
contain terms
proportional to $(D_\mu\varphi)\gamma^\mu\ol\epsilon$
with $\varphi=\sigma$ and $\phi$, respectively.
$\delta^0\ol\lambda$ in 3d and $\delta^0\ol\psi$ also contain
similar scalar derivative terms.
In $d$-dimensional spacetime,
we can covariantize terms of this form by the replacement
\begin{align}
(D_\mu\varphi)
\gamma^\mu\ol\epsilon
\rightarrow
&(D_\mu\varphi)
\gamma^\mu\ol\epsilon
+\frac{2\Delta_\varphi}{d}\varphi\gamma^\mu D_\mu\ol\epsilon,
\nonumber\\
(D_\mu\varphi)
\gamma^\mu\epsilon
\rightarrow
&(D_\mu\varphi)
\gamma^\mu\epsilon
+\frac{2\Delta_\varphi}{d}\varphi\gamma^\mu D_\mu\epsilon.
\label{covdf}
\end{align}
The fermion derivative terms
in $\delta^0 D$ $\delta^0F$, and $\delta^0F^\dagger$
are covariantized by the replacement
\begin{align}
(\ol\epsilon\gamma^\mu D_\mu\chi)\rightarrow
&(\ol\epsilon\gamma^\mu D_\mu\chi)
+\frac{2\Delta_\chi+1-d}{d}(D_\mu\ol\epsilon\gamma^\mu\chi),
\nonumber\\
(\epsilon\gamma^\mu D_\mu\ol\chi)\rightarrow
&(\epsilon\gamma^\mu D_\mu\ol\chi)
+\frac{2\Delta_\chi+1-d}{d}(D_\mu\epsilon\gamma^\mu\ol\chi).
\label{covariantization}
\end{align}
We can easily confirm that (\ref{covdf}) and (\ref{covariantization})
are transformed covariantly by the Weyl transformation
(\ref{weyle}) and
(\ref{weylphi}) as fields with weight
$\Delta_\varphi+1/2$ and
$\Delta_\chi+1/2$, respectively.

\section{${\bf S}^3$ partition function}\label{partition.sec}
In this section we briefly review the computation
of the ${\bf S}^3$ partition function.
We here only consider the case with $\mu_I=\zeta_A=k_a=s=0$.

Both a 3d ${\cal N}=2$ theory
and a 4d ${\cal N}=1$ theory have
eight supercharges.
Four of them ($Q$ and $\ol S$)
correspond to the parameter $\epsilon$
and the other four ($\ol Q$ and $S$) to $\ol\epsilon$.
When we use localization, we choose
a nilpotent supercharge ${\cal Q}$,
and add ${\cal Q}$-exact terms
to the action.
Because we should use a linear combination
of $\ol Q$ and $S$ for computation of the index
(\ref{indexdef}),
we consider only transformations by $\ol\epsilon$
in the following.

On a conformally flat 3d background
the parameter $\ol\epsilon$ must satisfy the Killing equation\cite{Kapustin:2009kz}
\begin{equation}
D_m\ol\epsilon=\gamma_m\kappa,
\label{killing}
\end{equation}
where $\kappa$ is an arbitrary spinor.
Corresponding to four supercharges $\ol Q_a$ and $S_a$,
there are four linearly independent solutions to (\ref{killing}).
In the case of ${\bf S}^3$,
two of them are right-invariant, and belong to the $({\bf 2},{\bf 1})$ representation
of the $SO(4)=SU(2)_L\times SU(2)_R$ isometry group.
Let us denote spinors with $J_L=+1/2$ and $J_L=-1/2$ by
$\ol\epsilon_1$ and $\ol\epsilon_2$, respectively.
We adopt $\delta(\ol\epsilon_1)$ as ${\cal Q}$.
Both $\ol\epsilon_1$ and $\ol\epsilon_2$ satisfy
\begin{equation}
D_m\ol\epsilon
=-\frac{i}{2r}\gamma_m\ol\epsilon,
\label{leftkilling}
\end{equation}
and $(\ol\epsilon_1\ol\epsilon_2)$ is constant on ${\bf S}^3$.
The other two solutions of
(\ref{killing}), which we will not use in this paper,
are left-invariant, and
satisfy a similar equation to
(\ref{leftkilling})
with opposite sign on the right hand side.
In the following the parameter $\ol\epsilon$ is always
assumed to satisfy (\ref{leftkilling}).

The $\delta(\ol\epsilon)$ transformation laws for fields
on ${\bf S}^3$ are
obtained from (\ref{d0vector}) and (\ref{d0chiral}) by the Weyl-covariantization.
The vector multiplet transformation laws are
\begin{align}
&\delta(\ol\epsilon) A_m=-i(\ol\epsilon\gamma_m\lambda),\quad
\delta(\ol\epsilon) \sigma=(\ol\epsilon\lambda),\quad
\delta(\ol\epsilon) \lambda=0,\nonumber\\
&\delta(\ol\epsilon) \ol\lambda=
-\frac{i}{2}\gamma^{mn}\ol\epsilon F_{mn}
-\gamma^m\ol\epsilon D_m\sigma+D\ol\epsilon+\frac{i}{r}\ol\epsilon \sigma
,\nonumber\\
&\delta(\ol\epsilon)D=
-(\ol\epsilon\gamma^mD_m\lambda)
-(\ol\epsilon[\sigma,\lambda])
-\frac{i}{2r}(\lambda\ol\epsilon).
\label{d3vecsusy}
\end{align}
The chiral multiplet transformation laws are
\begin{align}
&\delta(\ol\epsilon) \phi^\dagger=\sqrt2(\ol\epsilon\ol\psi),\quad
\delta(\ol\epsilon) \phi=0,\quad
\delta(\ol\epsilon)\ol\psi=\sqrt{2}\ol\epsilon F^\dagger,\quad
\delta(\ol\epsilon) F^\dagger=0,\nonumber\\
&\delta(\ol\epsilon)\psi=
-\sqrt{2}\gamma^m\ol\epsilon D_m\phi
+\sqrt{2}\ol\epsilon\sigma\phi
+\frac{\sqrt2i}{r}\Delta_\Phi\ol\epsilon\phi
,\nonumber\\
&\delta(\ol\epsilon) F=
-\sqrt{2}(\ol\epsilon\gamma^mD_m\psi)
-\sqrt{2}\sigma(\ol\epsilon\psi)
-2(\ol\epsilon\ol\lambda)\phi
-\frac{\sqrt2i}{r}\left(\Delta_\Phi-\frac{1}{2}\right)(\ol\epsilon\psi),
\label{d3chisusy}
\end{align}
where $\Delta_\Phi$ is the Weyl weight of the chiral multiplet,
which is defined as the Weyl weight of the dynamical scalar component field.

There is an ambiguity in the choice of
the ${\cal Q}$-exact deformation Lagrangian density ${\cal L}$.
We adopt the following one obtained by applying
$\delta(\ol\epsilon_1)$ and
$\delta(\ol\epsilon_2)$ to an anti-chiral operator,
\begin{equation}
(\ol\epsilon_1\ol\epsilon_2){\cal L}=
\delta(\ol\epsilon_1)\delta(\ol\epsilon_2)
\left(-\frac{1}{4}\tr(\ol\lambda\ol\lambda)
-\frac{1}{2}\sum_I \phi_I^\dagger F_I\right),
\label{antichiral}
\end{equation}
where $\tr$ represents
a gauge invariant positive definite inner product.
(\ref{antichiral}) can be used both in 3d and 4d.
In 3d, by using the 3d transformation laws
(\ref{d3vecsusy}) and (\ref{d3chisusy}),
we obtain
\begin{align}
{\cal L}^{(3d)}
=&
\tr\bigg[
\frac{1}{4}F_{mn}F^{mn}
+\frac{1}{2}D_m \sigma D^m\sigma
-\frac{1}{2}\epsilon^{mnp}F_{mn}D_p\sigma
+\frac{1}{2}\left(\frac{1}{r}\sigma-iD\right)^2
\nonumber\\&
-(\ol\lambda\gamma^mD_m\lambda)
-(\ol\lambda[\sigma,\lambda])
+\frac{i}{2r}(\ol\lambda\lambda)\bigg]
\nonumber\\
&+
\sum_I\bigg[
-\phi_I^\dagger D_m D^m\phi_I
+\phi_I^\dagger\sigma\sigma\phi_I
+\phi_I^\dagger D\phi_I
\nonumber\\&
-\frac{i(1-2\Delta_I)}{r}\phi_I^\dagger\sigma\phi_I
-\frac{\Delta_I(\Delta_I-2)}{r^2}\phi_I^\dagger\phi_I
-F_I^\dagger F_I
\nonumber\\&
-(\ol\psi_I\gamma^mD_m\psi_I)
-(\ol\psi_I\sigma\psi_I)
-\frac{i(\Delta_I-\frac{1}{2})}{r}(\ol\psi_I\psi_I)
-\sqrt2\phi_I^\dagger(\lambda\psi_I)
-\sqrt2(\ol\psi_I\ol\lambda)\phi_I
\bigg].
\label{3dchirallag}
\end{align}
(We use notation that in Euclidean signature the Hermitian conjugate
of the auxiliary fields $D$ and $F_I$ are $-D$ and $-F_I^\dagger$, respectively.)
In the large $u$ limit,
we can perform the path integral (\ref{zs3}),
and obtain
the matrix model integral\cite{Kapustin:2009kz,Jafferis:2010un,Hama:2010av}
\begin{align}
Z=\int_{\RR^{\rank G}} d\sigma J^{(3d)}(\sigma)
Z^{\rm vector}(\sigma)\prod_IZ_{\Phi_I}^{\rm chiral}(\sigma).
\label{zresult}
\end{align}
Integration variable $\sigma$ in (\ref{zresult}) is
an element of the Cartan subalgebra of the gauge group $G$.
The Jacobian factor $J^{(3d)}(\sigma)$ is
\begin{equation}
J^{(3d)}(\sigma)=\prod_{\alpha\in \Delta}\pi\alpha(r\sigma).
\label{j3d}
\end{equation}
$Z^{\rm vector}(\sigma)$ and $Z^{\rm chiral}_{\Phi_I}(\sigma)$
are $1$-loop partition function of
vector and chiral multiplets.
They are given by
\begin{align}
Z^{\rm vector}(\sigma)
=&
\prod_{\alpha\in\Delta}\frac{\sinh(\pi\alpha(r \sigma))}{\pi\alpha(r \sigma)},
\nonumber\\
Z_{\Phi_I}^{\rm chiral}(\sigma)
=&
\prod_{\rho\in R_I}
\prod_{k=1}^\infty
\left(
\frac{k+1-\Delta_I-i\rho(r\sigma)}
{k-1+\Delta_I+i\rho(r\sigma)}
\right)^k.
\end{align}

\section{Superconformal index}\label{index.sec}
Let us consider a 4d ${\cal N}=1$ theory in ${\bf S}^3\times{\bf S}^1$.
The background is conformally flat, and the parameter $\ol\epsilon$ must satisfy
the Killing equation
\begin{equation}
D_\mu \ol\epsilon=\gamma_\mu\kappa.
\label{4dlikking}
\end{equation}
To relate 3d spinor $\ol\epsilon(x^m)$ satisfying
(\ref{leftkilling})
and 4d spinor $\ol\epsilon(x^\mu)$, we take the anzats
\begin{equation}
\ol\epsilon(x^\mu)=f(x^4)\ol\epsilon(x^m).
\end{equation}
From (\ref{leftkilling})
the 4d spinor $\ol\epsilon(x^\mu)$ satisfies
\begin{equation}
D_\mu\ol\epsilon(x^\mu)=\frac{1}{2r}\gamma_\mu\gamma_4\ol\epsilon(x^\mu)
\label{killingin4}
\end{equation}
for $\mu=1,2,3$.
For $\ol\epsilon$ to be a Killing spinor
in 4d, this must hold for $\mu=4$, too.
This determines the function $f(x^4)$ up to normalization as
\begin{equation}
f(x^4)= e^{\frac{x^4}{2r}}.
\end{equation}
Corresponding to the Killing spinors $\ol\epsilon_1(x^m)$ and $\ol\epsilon_2(x^m)$
in 3d, we define two Killing spinors in 4d, which
are denote by the same symbols $\ol\epsilon_1$ and $\ol\epsilon_2$.
We adopt $\delta(\ol\epsilon_1(x^\mu))$ as ${\cal Q}$ in the same way as in 3d.

We want to compute a quantity in the form
\begin{equation}
I=\tr[(-1)^F {\cal O}q^D],
\label{igen}
\end{equation}
where ${\cal O}$ is an operator constructed from the
Cartan generators of the superconformal and flavor symmetries.
The most general form of ${\cal O}$ is
\begin{equation}
{\cal O}=y^{-\frac{3}{2}R-2J_L}
t^{R+2J_L}
x^{2J_R}
\prod_i
h_i^{{\cal F}_i}.
\end{equation}
This is equivalent to
imposing the boundary condition
\begin{equation}
\Phi(x^m,x^4)={\cal O}
\Phi(x^m,x^4+\beta r),
\label{phibc}
\end{equation}
on an arbitrary field $\Phi$.
For localization to be applicable the supercharge ${\cal Q}$
must commute with the operator ${\cal O}$.
Equivalently, the Killing spinor $\ol\epsilon_1$ must
satisfy the boundary condition
(\ref{phibc}).
This requires
$y=q$, and in this case
(\ref{igen}) becomes the index (\ref{indexdef}).

The 4d supersymmetry transformation laws
are obtained from (\ref{d0vector}) and (\ref{d0chiral})
by using (\ref{covdf}), (\ref{covariantization}),
and (\ref{killingin4}).
The transformation laws for a vector multiplet are
\begin{align}
&\delta(\ol\epsilon) A_\mu=-i(\ol\epsilon\gamma_\mu\lambda),\quad
\delta(\ol\epsilon)\lambda=0,\nonumber\\
&\delta(\ol\epsilon)\ol\lambda=-\frac{i}{2}\gamma^{\mu\nu}\ol\epsilon F_{\mu\nu}+D\ol\epsilon,\nonumber\\
&\delta(\ol\epsilon) D=-(\ol\epsilon\gamma^\mu D_\mu\lambda).
\label{4dvector}
\end{align}
The chiral multiplet transformation laws are
\begin{align}
&\delta(\ol\epsilon) \phi^\dagger=\sqrt{2}(\ol\epsilon\ol\psi),\quad
\delta(\ol\epsilon) \phi=0,\quad
\delta(\ol\epsilon)\ol\psi=\sqrt{2}\ol\epsilon F^\dagger,\quad
\delta(\ol\epsilon) F^\dagger=0,\nonumber\\
&\delta(\ol\epsilon)\psi=
-\sqrt{2}\gamma^\mu\ol\epsilon D_\mu\phi
-\frac{\sqrt{2}\Delta_\Phi}{r}\gamma^4\ol\epsilon\phi,\nonumber\\
&\delta(\ol\epsilon) F
=-\sqrt{2}(\ol\epsilon\gamma^\mu D_\mu\psi)
-2(\ol\epsilon\ol\lambda)\phi
-\frac{\sqrt{2}(\Delta_\Phi-1)}{r}(\ol\epsilon\gamma^4\psi).
\label{4dchitr}
\end{align}

The 4d deformation Lagrangian density ${\cal L}^{(4d)}$
is given by (\ref{antichiral}) with
the 4d transformation laws (\ref{4dvector}) and
(\ref{4dchitr}),
\begin{align}
{\cal L}^{(4d)}=&\tr\bigg[
\frac{1}{4}F_{\mu\nu}F^{\mu\nu}
-\frac{1}{8}\epsilon^{\mu\nu\rho\sigma}F_{\mu\nu}F_{\rho\sigma}
-\frac{1}{2}D^2
-(\ol\lambda\gamma^\mu D_\mu\lambda)\bigg]
\nonumber\\
&+\sum_I\bigg[-F_I^\dagger F_I
   -\phi_I^\dagger D_\mu D^\mu \phi_I
   +\phi_I^\dagger D\phi_I
   -\frac{\Delta_I^2-2\Delta_I}{r^2}\phi_I^\dagger\phi_I
   -\frac{2(\Delta_I-1)}{r}\phi_I^\dagger D_4\phi_I
\nonumber\\&
   -(\ol\psi_I\gamma^\mu D_\mu\psi_I)
   -\frac{\Delta_I-1}{r}(\ol\psi_I\gamma^4\psi_I)
   -\sqrt2\phi_I^\dagger(\lambda\psi_I)
   -\sqrt2(\ol\psi_I\ol\lambda)\phi_I\bigg].
\label{4dlag}
\end{align}
Note that this action contains only the anti-self-dual part of $F_{\mu\nu}$,
and we need to change the coefficient of the topological term
$\propto\tr(F\wedge F)$ to localize the path integral to flat connections.
This is possible
because the index does not depend on the coefficient of this term
as well as other coupling constants consistent with the symmetry of the system.
This Lagrangian density is essentially the same as
what is
derived in \cite{Romelsberger:2005eg}.
The index can be computed exactly
by performing the path integral (\ref{ideform}) in the large $u$ limit.
The result is\cite{Romelsberger:2007ec}
\begin{equation}
I(t,x,h_i)=\int_{\TT^{\rank G}} dA_4 J^{(4d)}(A_4) \pexp f(q^{i r A_4},t,x,h_i).
\label{iresult}
\end{equation}
The $A_4$ integral is taken over the maximal torus of the gauge group $G$.
$\pexp$ is the plethystic exponential
\begin{equation}
\pexp f(g,t,x,h_i)=
\exp\left(\sum_{m=1}^\infty\frac{1}{m}f(g^m,t^m,x^m,h_i^m)\right).
\end{equation}
$J^{(4d)}(A_4)$ is the Jacobian factor associated with the gauge fixing,
\begin{equation}
J^{(4d)}(A_4)=\prod_{\alpha\in\Delta}\frac{\sin(\pi\beta\alpha( r A_4))}{\beta}.
\label{j4d}
\end{equation}
$f(g,t,x,h_i)$ is the letter index.
The contribution of vector multiplets
is
\begin{align}
&f^{\rm vector}(q^{ir A_4},t,x,h_i)
\nonumber\\
&=
\sum_{\alpha\in G}q^{i\alpha(r A_4)}\left(
\sum_{l=0}^\infty\sum_{k=-l/2}^{l/2}
t^{l+2}x^{2k}
-
\sum_{l=1}^\infty\sum_{k=-l/2}^{l/2}
t^lx^{-2k}
\right)
\nonumber\\
&=
\frac{2t^2-t(x+x^{-1})}
{(1-tx)(1-tx^{-1})}
\sum_{\alpha\in G}q^{i\alpha(r A_4)}.
\label{vectorletter}
\end{align}
The contribution of a chiral multiplet $\Phi_I$
belonging to a gauge representation $R_I$ is
\begin{align}
&f^{\rm chiral}_{\Phi_I}(q^{ir A_4},t,x,h_i)
\nonumber\\
&=
\sum_{\rho\in R_I}\sum_{l=0}^\infty
\sum_{k=-l/2}^{l/2}
\left(
q^{i\rho(r A_4)}t^{l+\frac{2}{3}\Delta_I}x^{-2k}\prod_ih_i^{{\cal F}_i(\Phi_I)}
-q^{-i\rho(r A_4)}t^{l-\frac{2}{3}\Delta_I+2}x^{2k}\prod_ih_i^{-{\cal F}_i(\Phi_I)}
\right)
\nonumber\\
&=
\sum_{\rho\in R_I}\frac{
q^{i\rho(r A_4)}
t^{\frac{2}{3}\Delta_I}
\prod_ih_i^{{\cal F}_i(\Phi_I)}
-
q^{-i\rho(r A_4)}
t^{2-\frac{2}{3}\Delta_I}
\prod_ih_i^{-{\cal F}_i(\Phi_I)}
}
{(1-tx)(1-tx^{-1})}.
\label{chiralletter}
\end{align}

\section{Comparison of the deformation actions}\label{comparison.sec}
In order to relate the ${\bf S}^3$ partition function and the index,
let us compare the Lagrangian densities ${\cal L}^{(3d)}$
in (\ref{3dchirallag}) and ${\cal L}^{(4d)}$ in (\ref{4dlag}).
They
look similar, but not the same.
The difference is partially absorbed by shifting the auxiliary $D$-field.
\begin{equation}
D^{(4d)}= D^{(3d)}+\frac{i}{r}\sigma.
\label{dshift}
\end{equation}
Even after this shift the actions are still different.
If we assume there are no non-trivial background Wilson lines
around ${\bf S}^1$
and the covariant derivative
$D_4$ reduces to $-iA_4=-i\sigma$ in dimensional reduction,
the difference is
\begin{equation}
{\cal L}^{(3d)}
-{\cal L}^{(4d)}
=\frac{1}{2r}\left[
(\ol\lambda\gamma^4\lambda)
-\sum_I(\ol\psi_I\gamma^4\psi_I)\right].
\label{additional}
\end{equation}
This difference can be removed by introducing
a suitable Wilson line
if the theory has the symmetry $R_0$ with the charge assignments
\begin{equation}
R_0(\lambda)=+1,\quad
R_0(\psi_I)=-1,\quad
R_0(A_\mu)=R_0(\phi)=0.
\end{equation}
We weakly gauge this symmetry and introduce
the gauge field $V_\mu$ for this symmetry.
If we turn on the Wilson line
\begin{equation}
\langle V_4\rangle=-\frac{i}{2r},
\end{equation}
the difference (\ref{additional})
is canceled by the terms arising from
the 4d fermion kinetic terms in (\ref{4dlag}).
This is equivalent to the insertion of the operator
\begin{equation}
{\cal O}=q^{-\frac{1}{2}R_0}
\end{equation}
in (\ref{igen}),
and thus we expect that the partition function $Z$
is given by
\begin{equation}
Z=\lim_{q\rightarrow 1}\tr[(-1)^Fq^{-\frac{1}{2}R_0}q^{D}].
\label{zlimtr}
\end{equation}

If the 4d parent theory
has non-vanishing superpotential,
the symmetry $R_0$ is in general broken.
However, the superpotential does not affect the index.
The relevant part of the deformed action $S_{\rm rel}^{(4d)}$
has the large symmetry
rotating chiral multiplets independently.
Let ${\cal F}_I$ denote the generator rotating only a chiral multiplet $\Phi_I$
by charge $1$.
Correspondingly, we introduce chemical potentials $h_I$.
The symmetry $R_0$ is related to $R$,
the $R$-symmetry in the superconformal algebra, by
\begin{equation}
R_0=R-\frac{2}{3}\sum_I\Delta_I{\cal F}_I,
\end{equation}
and we can express (\ref{zlimtr}) as a special limit of
the index,
\begin{equation}
Z=\lim_{q\rightarrow 1} I(t=q,x=1,h_I=q^{\frac{1}{3}\Delta_I}).
\label{relation0}
\end{equation}

It is easily checked that
this relation indeed holds for
(\ref{zresult}) and (\ref{iresult}) as follows.
Because the radius of the maximal torus $\TT^{\rank G}$ is inversely proportional
to the ${\bf S}^1$ period $\beta r$, it becomes
$\RR^{\rank G}$ in the limit $\beta\rightarrow 0$.
We also see that the Jacobian factor (\ref{j4d}) reduces to (\ref{j3d})
when $\beta\rightarrow 0$.
We obtain
\begin{equation}
\lim_{q\rightarrow 1}\int_{\TT^{\rank G}} dA_4J^{(4d)}(A_4)
=\int_{\RR^{\rank G}} d\sigma J^{(3d)}(A_4).
\label{intrel}
\end{equation}
For the letter indices,
we first express the plethystic exponential
of (\ref{vectorletter}) and (\ref{chiralletter})
as the infinite products,
\begin{align}
\pexp f^{\rm vector}(q^{ir A_4},t,x,h_I)
=&
\prod_{\alpha\in G}\frac{\prod_{l=1}^\infty\prod_{k=-l/2}^{l/2}
(1-q^{i\alpha(r A_4)}t^lx^{-2k})}
{
\prod_{l=0}^\infty\prod_{k=-l/2}^{l/2}
(1-q^{i\alpha(r A_4)}t^{l+2}x^{2k})},
\nonumber\\
\pexp f^{\rm chiral}_{\Phi_I}(q^{ir A_4},t,x,h_I)
=&
\prod_{\rho\in R_I}
\prod_{l=0}^\infty
\prod_{k=-l/2}^{l/2}
\left(
\frac{1-q^{-i\rho(r A_4)}t^{l-\frac{2}{3}\Delta_I+2}x^{2k}h_I^{-{\cal F}_I}}
{1-q^{i\rho(r A_4)}t^{l+\frac{2}{3}\Delta_I}x^{-2k}h_I^{{\cal F}_I}}
\right).
\label{pexpform}
\end{align}
Once we obtain these infinite products, it is straightforward to
confirm
the following relations.
\begin{align}
\lim_{q\rightarrow1}\pexp f^{\rm vector}(q^{ir A_4},q,1,q^{\frac{1}{3}\Delta_I})=&Z^{\rm vector}(A_4),
\nonumber\\
\lim_{q\rightarrow 1}\pexp f_{\Phi_I}^{\rm chiral}(q^{ir A_4},q,1,q^{\frac{1}{3}\Delta_I})=&Z^{\rm chiral}_{\Phi_I}(A_4).
\label{letterlimit}
\end{align}
Combining (\ref{intrel}) and (\ref{letterlimit}),
we obtain the relation 
(\ref{relation0}).

Before ending this section,
let us argue the anomaly associated with the 
inserted operator ${\cal O}$.
We consider the quantity
\begin{equation}
I(t,x,h_I)=\tr[(-1)^F q^X q^D],
\label{xinserted}
\end{equation}
where we denote the inserted operator ${\cal O}$ by $q^X$.
As we mentioned in \S\ref{intro.sec} the symmetry
generated by $X$
may be anomalous,
and then
the quantity
(\ref{xinserted})
is not well defined.
This can be regarded as inconsistency in the ${\bf S}^1$ compactification.
If $X$ is anomalous, the rotation by $q^X$ does not keep
the effective action $\Gamma$ invariant but changes it by
\begin{align}
\Gamma\rightarrow \Gamma'
=\Gamma+\int_{{\bf S}^3\times{\bf S}^1}\frac{\beta}{8\pi^2}\tr_F(XF\wedge F)
\label{gammashify}
\end{align}
where $\tr_F$ is the trace over Weyl fermions of positive chirality, which
contribute to the anomaly.

This change of the effective action obstacles the compactification
$x^4+\beta r\sim x^4$.
We can remove this obstruction by adding the following term
to the tree-level action.
\begin{align}
S'
&=-\int_{{\bf S}^3\times{\bf S}^1}\frac{x^4}{8\pi^2 r} \tr_F(XF\wedge F)
\nonumber\\
&=\int_{{\bf S}^1}dx^4\int_{{\bf S}^3}\frac{1}{8\pi^2 r} \tr_F\left[X\left(A\wedge F-\frac{2i}{3}A\wedge A\wedge A\right)\right].
\label{improve}
\end{align}
Due to the $x^4$ dependence of the $\theta$ angle,
the change of $S'$ under the shift $x^4\rightarrow x^4+\beta r$
cancels the anomalous change (\ref{gammashify}).
With the inclusion of the term
(\ref{improve}) in the action,
we can consistently compactify the $x^4$ direction
with the twist by ${\cal O}=q^X$.
When $X$ is anomalous, we define the quantity
(\ref{xinserted}) by the path integral (\ref{indexdef})
with the action improved by (\ref{improve}).

Let us consider whether it is possible to extend the additional term $S'$
in a supersymmetric way.
(\ref{improve}) is a three-dimensional Chern-Simons term except that
fields depend on the fourth coordinate $x^4$ along ${\bf S}^1$.
If all fields were $x^4$-independent,
we could actually construct the supersymmetric completion
\begin{align}
S'_{\rm SUSY}
=&
\int_{{\bf S}^1\times{\bf S}^3}d^4x\frac{\sqrt{g}}{8\pi^2 r}
\nonumber\\&\tr_F
\left[X\left\{\frac{i}{2}\epsilon^{mnp}\left(A_m\partial_n A_p-\frac{2i}{3}A_mA_nA_p\right)
+(\lambda\ol\lambda)
-D^{(4d)}A_4
+\frac{i}{r}A_4^2\right\}
\right].
\label{lpsusy}
\end{align}
For fields depending on $x^4$, however,
this action is not supersymmetry invariant.
We have non-vanishing supersymmetry transformation of the action
\begin{equation}
\delta
S'_{\rm SUSY}=
\int_{{\bf S}^1\times{\bf S}^3}d^4x\frac{\sqrt{g}}{8\pi^2 r}
\tr_F[X(\lambda\gamma^\mu\ol\epsilon)\partial_4 A_\mu].
\end{equation}
It is even worse that
(\ref{lpsusy}) is not even gauge invariant
due to terms containing $A_4$.
Unfortunately, we could not remedy these defects in
(\ref{lpsusy}),
and we use the non-supersymmetric term (\ref{improve}) to
turn on non-trivial Wilson lines for
anomalous symmetries.
In the large $u$ limit, the term
(\ref{improve}) is irrelevant, and
$I(t,x,h_I)$ is still given by the formula (\ref{iresult}).
However, the absence of the supersymmetry spoils the
$u$-independence of the path integrals, and
we can no longer regard $I(t,x,h_I)$ computed by the formula (\ref{iresult})
as the index of the original theory.
In the small ${\bf S}^1$ limit $\beta\rightarrow 0$, the term (\ref{improve}) vanishes
and the relation (\ref{relation0}) still holds.

\section{Generalization}\label{generalization.sec}
Up to here we have been assuming that
parameters of the 3d theory, $\mu_I$, $\zeta_A$, $k_a$, and $s$ all vanish.
Let us consider how we can obtain partition function
for a theory with these parameters turned on.

If the 3d theory has a flavor $U(1)$ symmetry,
we can introduce a real mass proportional to
the flavor charge for each chiral multiplet.
We here focus only on the relevant part $S_{\rm rel}^{(3d)}$,
and we can introduce real mass $\mu_I$ for each chiral multiplet $\Phi_I$
by weakly gauging ${\cal F}_I$ and turning on the
scalar component $\sigma_I$ of the corresponding vector multiplet $(\sigma_I,A_{I,m},\lambda_I,\ol\lambda_I,D_I)$.
(If some of ${\cal F}_I$ are anomalous, we need to introduce
the term (\ref{improve}) in the definition of the index.)
Note that we should turn on the auxiliary field $D_I$, too, to preserve
the supersymmetry (\ref{d3vecsusy}).
\begin{equation}
\langle \sigma_I\rangle=\mu_I,\quad
\langle D^{(3d)}_I\rangle=-\frac{i}{r}\mu_I,\quad
\langle A_{I,m}\rangle=
\langle\lambda_I\rangle=
\langle\ol\lambda_I\rangle=0.
\end{equation}
From the viewpoint of 4d theory, this is realized by
turning on the Wilson line for the flavor symmetry ${\cal F}_I$,
\begin{equation}
\langle A_{I,4}\rangle=\mu_I,\quad
\langle D^{(4d)}_I\rangle=
\langle A_{I,m}\rangle=
\langle\lambda_I\rangle=
\langle\ol\lambda_I\rangle=0.
\end{equation}
This is equivalent to the insertion of the operator
\begin{equation}
q^{-ir\sum_I\mu_I{\cal F}_I}
\label{fiinsertion}
\end{equation}
in (\ref{zlimtr}).

The next parameter we consider is a squashing parameter $s$.
The partition function of a theory on squashed ${\bf S}^3$ is
investigated in \cite{Hama:2011ea},
and it is found that
the partition function is changed
when both the isometries $SU(2)_L$ and $SU(2)_R$
are broken to $U(1)$.
It is proposed recently in \cite{Gadde:2011ia}
that the partition function depending on the squashing parameter
is reproduced from the index by turning on
$SU(2)_R$ Wilson line
in the case of 4d ${\cal N}=2$ theories.
We consider the insertion of the operator
\begin{equation}
q^{2sJ_R},
\label{jrinsertion}
\end{equation}
in a general 4d ${\cal N}=1$ theory.

By inserting (\ref{fiinsertion}) and (\ref{jrinsertion}) into
(\ref{zlimtr}),
we obtain
\begin{align}
Z=&\lim_{q\rightarrow 1}\tr[(-1)^Fq^{-\frac{1}{2}R_0}q^{-ir\mu_I{\cal F}_I}q^{2sJ_R}q^D]
\nonumber\\
=&\lim_{q\rightarrow 1}
I(t=q,x=q^s,h_I=q^{-ir\mu_I+\frac{1}{3}\Delta_I}).
\label{final}
\end{align}
This is the relation (\ref{mainresult})
with vanishing FI parameters.
Let us confirm that
(\ref{final}) reproduces
the partition function
of a 3d theory
with non-vanishing real masses and squashing parameter.
From the infinite product representation (\ref{pexpform})
we easily obtain
\begin{align}
&
\lim_{q\rightarrow 1}\pexp
f_{\Phi_I}^{\rm chiral}(q^{ir A_4},q,q^s,q^{r\mu_I+\frac{1}{3}\Delta_I})
\nonumber\\
&=
\prod_{\rho\in R}
\prod_{m,n\geq 0}
\left(
\frac{m(1+s)+n(1-s)-\Delta_I+2-i\rho(r A_4)-ir\mu_I}
{m(1-s)+n(1+s)+\Delta_I+i\rho(r A_4)+ir\mu_I}
\right),\nonumber\\
&\lim_{q\rightarrow 1}\pexp
f^{\rm vector}(q^{ir A_4},q,q^s,q^{r\mu_I+\frac{1}{3}\Delta_I})
\nonumber\\
&=
\prod_{\alpha\in G}\frac{\prod_{m,n\geq 0,(m,n)\neq(0,0)}
(m(1-s)+n(1+s)+i\alpha(r A_4))}
{
\prod_{m,n\geq0}
(m(1+s)+n(1-s)+2+i\alpha(r A_4)}.
\label{dblsine}
\end{align}
These are consistent with known results.
When $\mu_I=0$, these agree with the results
in \cite{Hama:2011ea} by the identification of parameters
\begin{equation}
\frac{\wt\ell}{\ell}=\frac{1+s}{1-s}.
\end{equation}
The $\mu_I$ dependence of
(\ref{dblsine})
is consistent with
the holomorphic dependence of the
partition function on $\Delta_I+ir\mu_I$\cite{Jafferis:2010un}.

One may think that this result is inconsistent with the result
in \cite{Hama:2011ea} because
the expression (\ref{final}) for the partition function
does not break the $SU(2)_L$ symmetry.
Ref \cite{Hama:2011ea} shows that an $SU(2)\times U(1)$ invariant
squashing does not change the partition function.
The reason for these different results is as follows.
The squashing considered in \cite{Hama:2011ea} is a left-invariant
squashing which preserves $SU(2)_L$ isometry,
and a Wilson line is turned on so that a half of
left-invariant Killing spinors is preserved.
There is in fact another essentially inequivalent possibility.
We can realize a left-invariant squashing with right-invariant Killing spinors
by taking a different
graviphoton background
from \cite{Hama:2011ea}.
In the above we use right-invariant Killing spinors,
and the $SU(2)_R$ Wilson line (\ref{jrinsertion})
preserves $SU(2)_L$ isometry.
This is a different situation from \cite{Hama:2011ea}.

In our case, the squashed metric is obtained from
the 4d background metric corresponding to
the insertion (\ref{jrinsertion})
\begin{align}
ds^2
=&
r^2\left[
(\mu^1)^2+(\mu^2)^2+(\mu^3+is dx^4)^2
+(dx^4)^2
\right]
\nonumber\\
=&r^2\left[
(\mu^1)^2+(\mu^2)^2
+\frac{1}{1-s^2}(\mu^3)^2
+(1-s^2)\left(dx^4
+\frac{is}{1-s^2}\mu^3\right)^2
\right],
\end{align}
where $\mu^a$ are left-invariant one-forms used in \cite{Hama:2011ea}.
We can read off the squashed metric of the base manifold,
\begin{equation}
ds^2=r^2[(\mu^1)^2+(\mu^2)^2]
+\frac{r^2}{1-s^2}(\mu^3)^2.
\end{equation}
It is interesting problem to confirm directly in 3d
that the partition function
for this squashed manifold with right-invariant Killing spinors
agree with (\ref{dblsine}).

As the last extension, let us introduce FI parameters.
Let $(A_{A,m},\sigma_A,\lambda_A,\ol\lambda_A,D_A^{(3d)})$
be $U(1)$ vector multiplets
for which we want to turn on the FI parameters.
If the 3d original action
contains the supersymmetry completion of FI terms
\begin{equation}
S^{(3d)}_{\rm FI}=-\sum_A\zeta_A^{(3d)}\int_{{\bf S}^3}\sqrt{g}\left(D^{(3d)}_A-\frac{i}{r}\sigma_A\right)d^3x,
\label{fi3d}
\end{equation}
the additional factor
\begin{equation}
\exp\left(-4\pi^2i r^2\sum_A\zeta^{(3d)}_A\sigma_A\right)
\label{fifactor}
\end{equation}
should be included
in the integrand in (\ref{zresult}).
$S_{\rm FI}^{(3d)}$ in (\ref{fi3d})
is obtained by dimensional reduction of 4d FI terms.
Note that the 4d FI term must be accompanied by smeared Wilson line
to preserve the supersymmetry,
\begin{equation}
S^{(4d)}_{\rm FI}=-\sum_A\zeta_A^{(4d)}\int_{{\bf S}^3\times{\bf S}^1}
\sqrt{g}\left(D^{(4d)}_A-\frac{2i}{r}A_{A,4}\right)d^4x.
\label{4dfi}
\end{equation}
Due to the coupling to the gauge fields, the 4d FI parameters
must be quantized, and thus the index can depend on them.
If we keep the relation $\beta r\zeta_A^{(4d)}=\zeta_A^{(3d)}$ in the small radius limit,
we reproduce 
(\ref{fi3d}) from
(\ref{4dfi}) and the factor
corresponding to (\ref{fifactor}) arises in
the index formula (\ref{iresult}).
Taking account of this relation,
we obtain the most general relation (\ref{mainresult}).

Finally we comment on Chern-Simons terms.
The supersymmetric completion of Chern-Simons term is
\begin{equation}
S_{\rm CS}^{(3d)}
=\int_{{\bf S}^3}\sqrt{g}\tr'\left[\frac{i}{2}\epsilon^{mnp}\left(A_m\partial_n A_p-\frac{2i}{3}A_mA_nA_p\right)
+(\lambda\ol\lambda)
-D\sigma\right]d^3x,
\label{ch3d}
\end{equation}
where $\tr'$ is a gauge invariant inner product containing Chern-Simons levels.
If these terms exist in the original action in (\ref{zs3}),
the extra factor
\begin{equation}
e^{-2\pi^2i\tr'(r^2\sigma^2)}
\end{equation}
arises in the integrand in (\ref{zresult}).
Unfortunately, we cannot reproduce this contribution from the
index due to the difficulty in constructing
4d action which gives Chern-Simons terms through dimensional reduction.

\section{Conclusions}\label{conc.sec}
In this paper we investigated
a relation between 3d and 4d actions used for computation of two
exactly calculable quantities,
the ${\bf S}^3$ partition function and
the 4d superconformal index.

When the 3d theory does not have Chern-Simons terms,
the relevant part of the action, which affects the ${\bf S}^3$
partition function,
consists of ${\cal Q}$-exact deformation terms
and the supersymmetric completion of FI terms.
In the case of round ${\bf S}^3$, we showed that this relevant part of the
3d action is obtained by dimensional reduction
from the corresponding terms in 4d action used for the computation
of the 4d superconformal index.
From this fact, we obtained a relation
which gives the ${\bf S}^3$ partition function as a
small radius limit of the 4d superconformal index
suitably generalized so that we can introduce chemical potentials
to anomalous symmetries.

To obtain the most general relation (\ref{mainresult}),
we used a connection between a squashing of ${\bf S}^3$
and $SU(2)_R$ Wilson line.
Although the squashing we considered in this paper,
the left-invariant squashing with right-invariant Killing spinors,
is different from squashings studied in \cite{Hama:2011ea},
our result agree with the partition function
for the $U(1)\times U(1)$ symmetric squashed ${\bf S}^3$ derived in \cite{Hama:2011ea}.

For 3d theory with Chern-Simons terms,
we could not give a 4d action reproducing the ${\bf S}^3$
partition function.

\section*{Acknowledgments}
I would like to thank Daisuke Yokoyama and Shuichi Yokoyama
for daily discussions.
I would also like to thank Kazuo Hosomichi and Yu Nakayama for
valuable comments.



\begin{thebibliography}{99}
\bibitem{Kapustin:2009kz}
  A.~Kapustin, B.~Willett and I.~Yaakov,
  ``Exact Results for Wilson Loops in Superconformal Chern-Simons Theories with
  Matter,''
  JHEP {\bf 1003}, 089 (2010)
  [arXiv:0909.4559 [hep-th]].
\bibitem{Jafferis:2010un}
  D.~L.~Jafferis,
  ``The Exact Superconformal R-Symmetry Extremizes Z,''
  arXiv:1012.3210 [hep-th].
\bibitem{Hama:2010av}
  N.~Hama, K.~Hosomichi and S.~Lee,
  ``Notes on SUSY Gauge Theories on Three-Sphere,''
  arXiv:1012.3512 [hep-th].
\bibitem{Kapustin:2010xq}
  A.~Kapustin, B.~Willett and I.~Yaakov,
  ``Nonperturbative Tests of Three-Dimensional Dualities,''
  JHEP {\bf 1010}, 013 (2010)
  [arXiv:1003.5694 [hep-th]].
\bibitem{Jafferis:2011ns}
  D.~Jafferis, X.~Yin,
  ``A Duality Appetizer,''
  [arXiv:1103.5700 [hep-th]].
\bibitem{Kapustin:2011gh}
  A.~Kapustin,
  ``Seiberg-like duality in three dimensions for orthogonal gauge groups,''
  arXiv:1104.0466 [hep-th].
\bibitem{Willett:2011gp}
  B.~Willett and I.~Yaakov,
  ``N=2 Dualities and Z Extremization in Three Dimensions,''
  arXiv:1104.0487 [hep-th].
\bibitem{Drukker:2010nc}
  N.~Drukker, M.~Marino and P.~Putrov,
  ``From weak to strong coupling in ABJM theory,''
  arXiv:1007.3837 [hep-th].
\bibitem{Herzog:2010hf}
  C.~P.~Herzog, I.~R.~Klebanov, S.~S.~Pufu and T.~Tesileanu,
  ``Multi-Matrix Models and Tri-Sasaki Einstein Spaces,''
  Phys.\ Rev.\  D {\bf 83}, 046001 (2011)
  [arXiv:1011.5487 [hep-th]].
\bibitem{Martelli:2011qj}
  D.~Martelli and J.~Sparks,
  ``The large N limit of quiver matrix models and Sasaki-Einstein manifolds,''
  arXiv:1102.5289 [hep-th].
\bibitem{Cheon:2011vi}
  S.~Cheon, H.~Kim and N.~Kim,
  ``Calculating the partition function of N=2 Gauge theories on $S^3$ and
  AdS/CFT correspondence,''
  arXiv:1102.5565 [hep-th].
\bibitem{Jafferis:2011zi}
  D.~L.~Jafferis, I.~R.~Klebanov, S.~S.~Pufu, B.~R.~Safdi,
  ``Towards the F-Theorem: N=2 Field Theories on the Three-Sphere,''
  [arXiv:1103.1181 [hep-th]].
\bibitem{Kinney:2005ej}
  J.~Kinney, J.~M.~Maldacena, S.~Minwalla and S.~Raju,
  ``An Index for 4 dimensional super conformal theories,''
  Commun.\ Math.\ Phys.\  {\bf 275}, 209 (2007)
  [arXiv:hep-th/0510251].
\bibitem{Romelsberger:2005eg}
  C.~Romelsberger,
  ``Counting chiral primaries in N = 1, d=4 superconformal field theories,''
  Nucl.\ Phys.\  B {\bf 747}, 329 (2006)
  [arXiv:hep-th/0510060].
\bibitem{Romelsberger:2007ec}
  C.~Romelsberger,
  ``Calculating the Superconformal Index and Seiberg Duality,''
  arXiv:0707.3702 [hep-th].
\bibitem{Dolan:2008qi}
  F.~A.~Dolan, H.~Osborn,
  ``Applications of the Superconformal Index for Protected Operators and q-Hypergeometric Identities to N=1 Dual Theories,''
  Nucl.\ Phys.\  {\bf B818}, 137-178 (2009).
  [arXiv:0801.4947 [hep-th]].
\bibitem{Spiridonov:2008zr}
  V.~P.~Spiridonov, G.~S.~Vartanov,
  ``Superconformal indices for N = 1 theories with multiple duals,''
  Nucl.\ Phys.\  {\bf B824}, 192-216 (2010).
  [arXiv:0811.1909 [hep-th]].
\bibitem{Spiridonov:2009za}
  V.~P.~Spiridonov, G.~S.~Vartanov,
  ``Elliptic hypergeometry of supersymmetric dualities,''
  [arXiv:0910.5944 [hep-th]].
\bibitem{Nakayama:2005mf}
  Y.~Nakayama,
  ``Index for orbifold quiver gauge theories,''
  Phys.\ Lett.\  B {\bf 636}, 132 (2006)
  [arXiv:hep-th/0512280].
\bibitem{Nakayama:2006ur}
  Y.~Nakayama,
  ``Index for supergravity on AdS(5) x T**(1,1) and conifold gauge theory,''
  Nucl.\ Phys.\  B {\bf 755}, 295 (2006)
  [arXiv:hep-th/0602284].
\bibitem{Benvenuti:2006qr}
  S.~Benvenuti, B.~Feng, A.~Hanany and Y.~H.~He,
  ``Counting BPS operators in gauge theories: Quivers, syzygies and
  plethystics,''
  JHEP {\bf 0711}, 050 (2007)
  [arXiv:hep-th/0608050].
\bibitem{Gadde:2010en}
  A.~Gadde, L.~Rastelli, S.~S.~Razamat and W.~Yan,
  ``On the Superconformal Index of N=1 IR Fixed Points: A Holographic Check,''
  arXiv:1011.5278 [hep-th].
\bibitem{Dolan:2011rp}
  F.~A.~H.~Dolan, V.~P.~Spiridonov, G.~S.~Vartanov,
  ``From 4d superconformal indices to 3d partition functions,''
  [arXiv:1104.1787 [hep-th]].
\bibitem{Gadde:2011ia}
  A.~Gadde and W.~Yan,
  ``Reducing the 4d Index to the $S^3$ Partition Function,''
  arXiv:1104.2592 [hep-th].
\bibitem{Hama:2011ea}
  N.~Hama, K.~Hosomichi, S.~Lee,
  ``SUSY Gauge Theories on Squashed Three-Spheres,''
  [arXiv:1102.4716 [hep-th]].
\end{thebibliography}
\end{document}